\documentclass[12pt,english,preprint,superscriptaddress,showpacs]{revtex4-1}
\usepackage[T1]{fontenc}
\usepackage[latin9]{inputenc}
\usepackage{amsmath}
\usepackage{graphicx}
\usepackage{amssymb}

\makeatletter
\@ifundefined{textcolor}{}
{%
 \definecolor{BLACK}{gray}{0}
 \definecolor{WHITE}{gray}{1}
 \definecolor{RED}{rgb}{1,0,0}
 \definecolor{GREEN}{rgb}{0,1,0}
 \definecolor{BLUE}{rgb}{0,0,1}
 \definecolor{CYAN}{cmyk}{1,0,0,0}
 \definecolor{MAGENTA}{cmyk}{0,1,0,0}
 \definecolor{YELLOW}{cmyk}{0,0,1,0}
 }

\makeatother

\usepackage{babel}

\begin{document}

\title {The impact of spatial correlation on the tunneling dynamics of few-boson mixtures in a combined triple well and harmonic trap}

\author{Lushuai Cao}
\email{lcao@physnet.uni-hamburg.de}
\affiliation{Zentrum f\"{u}r Optische Quantentechnologien, Universit\"{a}t Hamburg, Luruper Chaussee 149, D-22761 Hamburg, Germany}
\author{Ioannis Brouzos}
\email{ibrouzos@physnet.uni-hamburg.de}
\affiliation{Zentrum f\"{u}r Optische Quantentechnologien, Universit\"{a}t Hamburg, Luruper Chaussee 149, D-22761 Hamburg, Germany}
\author{Budhaditya Chatterjee}
\email{bchatter@physnet.uni-hamburg.de}
\affiliation{Zentrum f\"{u}r Optische Quantentechnologien, Universit\"{a}t Hamburg, Luruper Chaussee 149, D-22761 Hamburg, Germany}
\author{Peter Schmelcher}
\email{pschmelc@physnet.uni-hamburg.de}
\affiliation{Zentrum f\"{u}r Optische Quantentechnologien, Universit\"{a}t Hamburg, Luruper Chaussee 149, D-22761 Hamburg, Germany}

\begin{abstract}
We investigate the tunneling properties of a two-species few-boson mixture in a one-dimensional triple well and harmonic trap. The mixture is prepared in an initial state with a strong spatial correlation for one species and a complete localization for the other species. We observe a correlation-induced tunneling process in the weak interspecies interaction regime. The onset of the interspecies interaction disturbes the spatial correlation of one species and induces tunneling among the correlated wells. The corresponding tunneling properties can be controlled by the spatial correlations with an underlying mechanism which is inherently different from the well known resonant tunneling process. We also observe the correlated tunneling of both species in the intermediate interspecies interaction regime and the tunneling via higher band states for strong interactions.

\end{abstract}


\maketitle

\section{Introduction}
Ultracold atoms in optical lattices have manifested themselves as a powerful and flexible tool for the study of quantum systems \cite{review1,review2,review3}. Specially ultracold atoms in optical lattices possess a high degree of controllability. The geometry and strength of the optical lattice can be tuned by changing the frequency, amplitude and polarization of the involved lasers \cite{lattice-param0,lattice-param1,lattice-param2}, while the contact interaction strength of the atoms can be tuned via Feshbach resonances \cite{Feshbach-optical,Feshbach,Feshbach-magnetical} as well as through confinement induced resonances in quasi one-dimensional systems \cite{cir,CIR2,CIR3,CIR4,CIR5}. Moreover, their rich internal atomic properties offer the possibility to simulate a wide range of quantum systems, exploiting the atomic spin properties \cite{spinor-magnetism,spin-Hall,spin-orbital1,spin-orbital2} and the diversity of atomic species \cite{species-sascha,species-lm,species-sen,species-anderson,species-phaseseparation,species-counter,species-counter2,species-bfex,species-bf-thermal1,species-bf-thermal2,species-bf-thermal3}. This can lead to a plethora of quantum properties and phenomena which will deepen our understanding of quantum physics, as well as widen applications of quantum system.

Atomic species could differ in mass, intraspecies interaction strength, and in intraspecies permutation symmetry, $i.e.$ being bosons or fermions, which renders the study of mixed ultracold atomic systems highly attractive. In mixtures containing atoms of different masses, the heavier atoms can play the role of an effective potential for lighter atoms, and this has been theoretically predicted \cite{species-sascha,species-lm}, and also experimentally observed \cite{species-sen}. Such an effective potential can even lead to Anderson localization of both species \cite{species-anderson}. The interplay of intraspecies and interspecies interactions can give rise to a correlated phase structure, ranging from phase separation \cite{species-phaseseparation} to pair superfluid and counterflow superfluid \cite{species-counter,species-counter2}. In mixtures composed of bosons and fermions, various correlated phase structures have also been discovered \cite{species-bfex}, where the bosonic and fermionic species can be in dual Mott insulator phases, anti-correlated phases, as well as completely phase separated. Moreover, the interspecies interaction strength between the bosonic and fermionic species can reform the phase space, changing the temperature of the system, and induce an enhanced condensate of the bosonic species \cite{species-bf-thermal1,species-bf-thermal2,species-bf-thermal3}. The mixture of bosons and fermions also leaves an open question of how to explore the unique properties of boson-fermion mixtures, compared to boson-boson mixtures.

Quantum phase transitions represent a main focus of the study of cold atoms, and the theoretical prediction as well as the experimental observation of the superfluid and Mott-insulator phases manifested itself as a milestone of the field \cite{phase,phase-exp}. An interesting finding concerning the phase structure is the existence of domains composed of different phases, for instance, the domain structure of the Mott phase and superfluid phase in the optical lattice augmented by a harmonic trap \cite{domain0,domain-exp}. One of the important characteristics of such a domain structure are the spatial correlations between different lattice sites. Only the sites in the superfluid phase maintain nonzero correlations, while no correlations exist for the sites in the Mott phase. An open question concerning the domain structure is how the spatial correlations can affect the tunneling dynamics of the system.

In this work we study the effects of spatial correlations of a mixture of ultracold atoms on the quantum dynamics of the system. We do so by exploring a few-body bosonic mixture consisting of two bosonic species confined in a one-dimensional triple plus harmonic (TPH) well. We will firstly show that this few-body ensemble can present spatial correlations of the wavefunction between selected wells in the multiwell system, which is a signature of the domain structure in many body systems. We demonstrate how these spatial correlations can affect the tunneling dynamics of the bosonic mixture. In particular, it is shown that in the low interspecies interaction regime, this spatial correlation can induce tunneling between the correlated sites, which sheds a new light on the common resonant tunneling mechanism. Besides, in the intermediate interaction regime, we observe correlated tunneling of both species in resonant windows of interspecies interaction strength on top of delayed tunneling, and in the strong interaction regime, higher band states contribute to the tunneling, and enhanced tunneling via higher band states is found. 

This paper is organized as follows. In section 2, we discuss our model and first-principle numerical method the multi-configuration time-dependent Hartree (MCTDH) approach \cite{MCTDH1,MCTDH2,MCTDH3}. In section 3 we study the ground state evolution of the single-species bosons in the TPH well, and the occurrence of corresponding spatial correlations. Section 4 contains a study of the tunneling dynamics of a bosonic mixture in the TPH well for different interspecies interaction strengths. We distinguish three interspecies interaction regimes according to the tunneling behavior: the low interspecies interaction regime where correlation-induced tunneling dominates, the intermediate regime with correlated tunneling, and the high interaction regime, where higher band states come into play. Section 5 contains our summary.

\section{Setup and computational method}
\subsection{Setup}
Our aim is to study the tunneling dynamics of a bosonic mixture consisting of two bosonic species, A and B, in a TPH well. The Hamiltonian reads

\begin{equation}{
H=\sum_{\sigma=A,B}\sum_{j=1}^{N_{\sigma}}[-\frac{\hbar^{2}}{2M_{\sigma}}\partial_{x_{\sigma,j}}^{2}+V_{tr}(x_{\sigma,j})+\frac{1}{2}\sum_{k (k\neq j)}g_{\sigma,1D}\delta(x_{\sigma,j}-x_{\sigma,k})]+\sum_{i=1}^{N_{A}}\sum_{j=1}^{N_{B}}g_{AB,1D}\delta(x_{A,i}-x_{B,j})
}.
\end{equation}
Here, $V_{tr}(x_{\sigma})=V_{0}sin^{2}(\kappa x_{\sigma})+\frac{1}{2}M_{\sigma} \omega^{2}x_{\sigma}^{2}$ models the TPH trap, and we apply hard-wall boundary conditions at $x_{\sigma}=\pm 3\pi/2\kappa$ to confine the bosons to the central three adjacent wells. $M_{\sigma}$ is the mass of the $\sigma$-species, and $\kappa$ is the wave vector of the laser beams forming the optical lattice. We assume that the bosons of both species take the same mass, $i.e.$, $M_{A}=M_{B}\equiv M$.

We model the interspecies and intraspecies contact interaction by a delta function with respect to the relative coordinate of two bosons, and the strength of the interaction is given by $g_{AB,1D}$, $g_{A,1D}$ and $g_{B,1D}$ for the interspecies and the intraspecies interaction, respectively. In one-dimensional system as considered here, these interaction strengths can be tuned by Feshbach \cite{Feshbach-optical,Feshbach,Feshbach-magnetical} as well as confinement induced resonances \cite{CIR2,CIR3}. 

In the following we make use of the natural units $\hbar=1$,$M=1$, and $\kappa=1$, which is equivalent to rescaling the Hamiltonian in units of the recoil energy $E_{R}=\frac{\hbar^{2}\kappa^{2}}{2M}$. The competing parameters, which we focus on, are mainly the inter- and intraspecies interaction strength, $g_{AB}=g_{AB,1D}/E_{R}$ and $g_{\sigma}=g_{\sigma,1D}/E_{R}$, $\sigma=A,B$.

We aim to study the effects of spatial correlation on the tunneling dynamics. To identify the relevant mechanisms, we focus on a few-body mixture consisting of two bosons of A species, and one of B species. The initial state is prepared as follows: firstly the interspecies interaction $g_{AB}$ is turned off, and the A bosons relax to the ground state, which is characterized by an occupation of the middle well and spatial correlations between the occupied left and right wells, while the B boson is localized in the left well. To trigger the tunneling, we turn on the interspecies interaction, and simultaneously set the B boson free to move, which would drive the system out of its initial state, and initiates the tunneling process. In experiments, this setup could be realized by the sudden doping of a B boson into the left well of a TPH trap, where two A bosons have already relaxed to the ground state with spatial correlations between the left and right wells. Such an instantaneous doping is then accompanied by the release of the B boson to the TPH well as well as the onset of the interspecies interaction. The doping of the B boson can be experimentally achieved by $e.g.$ single-site addressability techniques, which are now available for ultracold atoms \cite{singlesite1,singlesite2}.

\subsection{Computational approach: the Multi-Configuration Time-Dependent Hartree Method}

In our study, we apply the numerically exact multi-configuration time-dependent Hartree method (MCTDH) \cite{MCTDH1,MCTDH2,MCTDH3}, which is a wave packet dynamical approach to the $\textit{ab initio}$ solution of multi-dimensional time-dependent Schr\"{o}dinger problems. In MCTDH, the many-body wave function is expanded in terms of Hartree products of single-particle functions of a corresponding basis:

\begin{equation}{
\Psi(x_{1},...,x_{N};t)=\sum_{j_{1},...,j_{N}}A_{j_{1},...,j_{N}}(t)\varphi^{(1)}_{j_{1}}(x_{1},t)...\varphi^{(N)}_{j_{N}}(x_{N},t)},
\end{equation}
where ${\varphi^{(i)}_{j_{i}}(x_{i})}$ are the single particle functions for the degree of freedom $x_{i}$. Applying the Dirac-Frenkel variation principle to this ansatz, leads to a set of differential equations for the A coefficients and the single particle functions, which provide us with the time evolution of the system. In this ansatz, both the expansion coefficients $A_{j_{1},...,j_{N}}$ and the single particle functions ${\varphi^{(i)}_{j_{i}}(x_{i})}$ are time-dependent, and due to the time dependence of ${\varphi^{(i)}_{j_{i}}}$, It is optimized at each time step, which leads to a reduction of the total number of necessary single-particle functions in order to achieve convergence, and manifest itself as a big advantage of MCTDH compared to other exact computational methods.

The multi-configuration expansion of the wave function in MCTDH intrinsically takes into account higher band effects, which is essential for the study of systems with a temporally varying Hamiltonian, where the system could be excited to higher band states. In our setup, the instantaneous onset of $g_{AB}$ can give rise to higher band excitation, beyond the single-band approximation, and therefore, renders MCTDH an appropriate tool for our investigation. As we study a bosonic mixtures, permutation symmetry within a single species of the mixture should also be preserved during propagation of the wave packet. This is done in MCTDH by symmetrizing the A-coefficients according to the permutation symmetry of each species, and using a single set of single particle wavefunctions for each species.

To prepare our initial condition in MCTDH, we choose the interspecies interaction strength $g_{AB}=0$, and we take the intraspecies interaction $g_{A}$, such that the ground state of the two A bosons exhibit spatial correlations only between the left and right wells. The localization of the B boson in the left well is realized by applying an artificial hard wall boundary condition between the left and middle wells for the B-species boson only.

\section{Ground state evolution of single species in the triple well with harmonic trap}

An milestone of ultracold atoms in optical lattices is the Mott insulator-superfluid phase transition \cite{phase,phase-exp} driven by the interaction strength, the hopping strength between neighboring sites and the chemical potential. In an optical lattice augmented by a harmonic trap, the atoms can also exhibit a domain structure consisting of different phases \cite{domain0,domain2,domain-exp}. This domain structure is characterized by the spatial correlation of the atomic wavefunction between selected lattice sites. For instance, one can realize a domain structure consisting of a Mott domain around the minimum of the harmonic confinement, surrounded by two superfluid wings. The superfluid wings are characterized by strong correlations, while there's no spatial correlation between different sites in the Mott domain. In this section we demonstrate that such spatially selective correlations can be also realized for few-body systems confined to a TPH trap.

 We start with two single-species bosons in a TPH well, and study the evolution of the ground state with increasing interaction strength. For the analysis of the ground state we focus on the evolution of the mean occupation $<n_{\alpha}>$ and particle fluctuation $\delta n_{\alpha}=<n^{2}_{\alpha}>-<n_{\alpha}>^{2}$ of each well, and the spatial correlations among different wells, defined as $G_{\alpha,\beta}=2Re(\langle \Psi|a_{\alpha}a^{+}_{\beta}|\Psi\rangle)$, where $\alpha$ and $\beta$ label the three wells, $|\Psi\rangle$ is the ground state, and $a_{\alpha}^{+}(a_{\alpha})$ is the creation (annihilation) operator on the well $\alpha$. The corresponding results for these quantities are shown in figure 1. Figure 1(a) presents the evolution of the mean occupation, and we observe a two-plateau behavior, where the mean occupation of the middle well changes from two to one, and that of the left and right wells changes from zero to $1/2$. Mean occupation of $1/2$ implies here that a boson occupies the left and right wells equally $i.e.$ with the same probability, leading to particle fluctuation in the left and right wells on the plateau II, as confirmed by figure 1(b). The middle well exhibits particle fluctuations only in the transition regime between plateaus I and II (see fig. 1(b)). In figure 1(c) the correlation function between different wells shows a nonzero correlation between the left and right wells on plateau II, and no correlations exist between the middle well with the others except in the transition regime between the plateaus. 

\begin{figure}
\includegraphics[width=16cm]{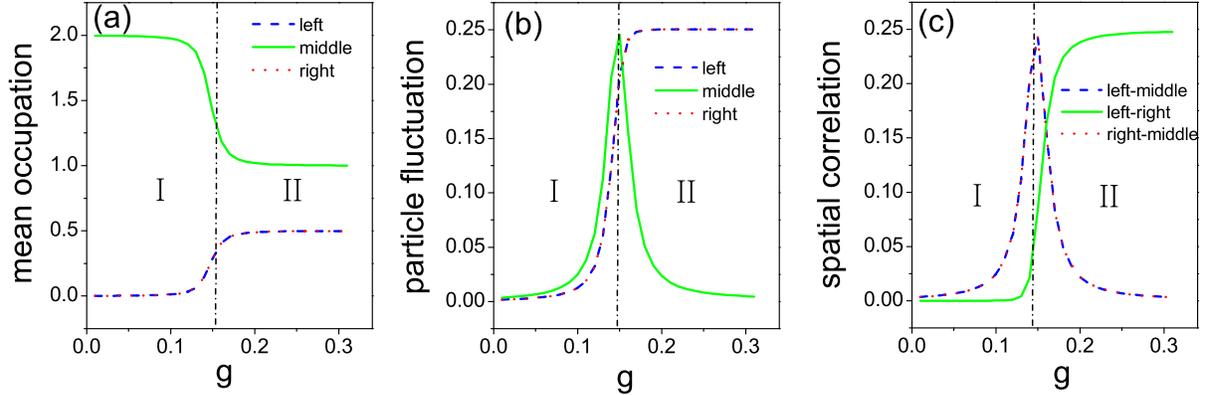}
\caption{The evolution of ground state of the system consisting of two A-species bosons in the TPH well, with respect to the interaction strength: (a) the mean occupation, (b) particle fluctuation, and (c) the spatial correlation between different wells. All the figures show a two-plateau structure, marked as plateau I and II, and this two-plateau structure corresponds to the transition of the ground state from $|0,2,0\rangle$ to $|1,1,0\rangle+|0,1,1\rangle$ (see discussion in text). The particle fluctuations and also the fluctuation induced correlations between the left and right wells arise on plateau II. The calculations are done by MCTDH, with a harmonic trap frequency of $\omega=0.01$.  } \label{figure 1}
\end{figure}

We use the number-state basis to analyze the results of figure 1. The number-state basis of two bosons in a triple well contains six number states: $\{|2,0,0\rangle$, $|1,1,0\rangle$, $|1,0,1\rangle$, $|0,2,0\rangle$, $|0,1,1\rangle$, $|0,0,2\rangle\}$, where $|n_{L},n_{M},n_{R}\rangle$ refers to the state with $n_{L}$,$n_{M}$, and $n_{R}$ bosons in the left, middle and right wells, respectively. When the interaction strength is low, all bosons concentrate in the middle well, corresponding to the ground state $|0,2,0\rangle$, and this refers to the plateau I in figure 1. As the interaction strength increases beyond some critical value, one boson is repelled out of the middle well, and localizes in the left and right wells simultaneously with the same probability, relating to the ground state of $|1,1,0\rangle+|0,1,1\rangle$, and consequently the mean occupation of the left and right well takes the value of $1/2$, while spatial correlations between the left and right wells arise. This regime corresponds to the second plateau II in figure 1. Only around the 'critical point' of the transition, the ground state possesses significant contributions from all the three number states $|0,2,0\rangle$, $|1,1,0\rangle$ and $|0,1,1\rangle$, and the middle well gets correlated with the left and right wells, which corresponds to the peaks of the occupation fluctuation of the middle well and the spatial correlation of the middle well with the other wells in figures 1(b) and (c), respectively.

\begin{figure}
\includegraphics[width=16cm]{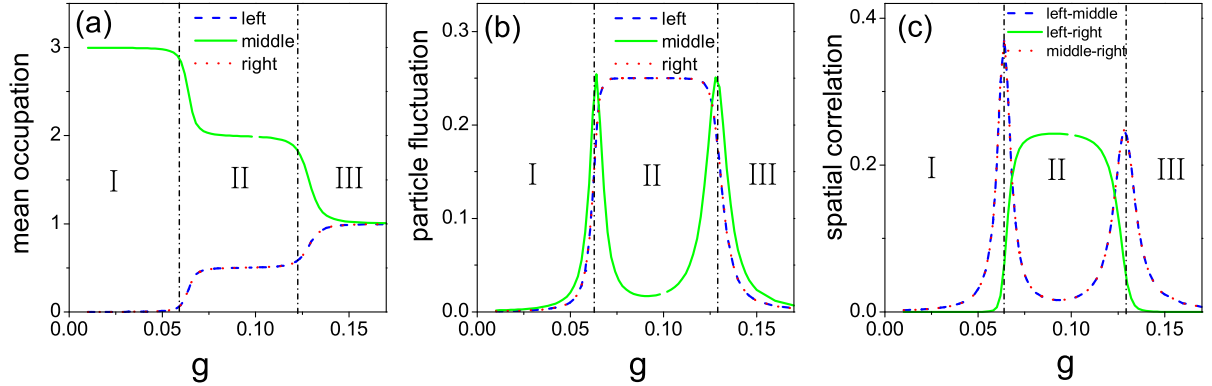}
\caption{The response of (a) mean occupation, (b) particle fluctuation, and (c) spatial correlation between different wells to the interaction strength, for the system of three A bosons in the TPH well. A three-plateau structure shows up and illustrates that the ground state undergoes transitions from $|0,3,0\rangle_{A}$ (plateau I) to $|1,2,0\rangle_{A}+|0,2,1\rangle_{A}$ (plateau II), and $|1,1,1\rangle_{A}$ (plateau III) at different interaction strengths. The parameters of the calculation are the same as in figure 1.  } \label{figure 2}
\end{figure}

In the system of two bosons in the TPH well we find the spatial site-selective correlations only between the left and right wells, and this system is the smallest system to realize such site-selective spatial correlations. In system with more bosons, such a multi-plateau structure of the ground state evolution and site-selective correlations are also encountered. Taking three bosons in the TPH well for example, we calculated the ground state evolution with respect to increasing interaction strength, and figure 2 shows the behavior of the mean occupation, particle fluctuation, as well as the spatial correlations. Similar to the case of two bosons, we observe a multi-plateau structure in the three domains, and finite spatial correlations of particular sites on the second plateau. Combing figure 1 and 2, it becomes clear that an increasing interaction strength drives the bosonic ensemble from localization to the central well to a spreading among several wells. During this spreading, bosons can occupy spatially symmetric wells simultaneously, and this gives rise to nonzero particle fluctuations and correlations between the related wells. This case corresponds to the formation of domain structures in many-body systems, and offers a microscopic explanation to the domain-structure formation in the harmonically dressed optical lattices.

\section{Tunneling dynamics in different interspecies interaction regime }

Let us now study the tunneling dynamics of a two-species bosonic ensemble, containing two bosons of A species and one of B species, which are confined to a one-dimensional TPH trap. Initially the interspecies interaction is turned off, and each species relaxes to the corresponding ground state. Following the discussion in section 3, we set the intraspecies interaction strength of the A species such as we realize the second plateau of figure 1, and the ground state of A bosons corresponds to spatial correlation only between the left and right wells. The single B particle is initially localized to the left well. We trigger the tunneling by releasing the B particle to the complete system, and simultaneously turning on the interspecies interaction $g_{AB}$ to a finite value. In this way the B particle will tunnel in the triple well on top of the domain-like structure of A species, while the tunneling of the B particle will also drive the A bosons out of equilibrium as a back action. We study the interplay between the spatial correlations of the A species and the tunneling of the B particle.

Generally speaking, we can divide the interspecies interaction into three regimes according to the tunneling properties. The first one is the low interspecies interaction regime, where the tunneling is dominated by the spatial correlations of the A species, and we observe the correlation-induced tunneling of the A species, while the B particle remains practically localized in its initial well. In the intermediate regime of $g_{AB}$, the correlation-induced tunneling vanishes due to a strong energetic detuning, and we observe a correlated tunneling process of A and B species in narrow windows of $g_{AB}$ on top of delayed tunneling. Finally, when the interspecies interaction is strong enough such that the system is non-adiabatically excited to higher bands during turning on $g_{AB}$, we observe an enhanced tunneling due to higher band contributions.

\subsection{Correlation-induced tunneling in the low interspecies interaction regime}

Firstly, we focus on the low interaction regime. The initial state can be expressed as $(|1,1,0\rangle_{A}+|0,1,1\rangle_{A})|1,0,0\rangle_{B}$, which indicates that the A bosons are in the ground state corresponding to the plateau II of figure 1(a), and the boson B is in the left well. We explore the time evolution of the population of the A and B bosons in each well, see figure 3. In figure 3(a), we show the evolution of the population of the A species, which oscillates coherently between the left and right wells, and remains almost constant in the middle well. The population of the B boson is shown in figure 1(b), and we find the B boson to be localized in the left well. The role of the B boson is an effective tilt in the left well. From figure 1 we conclude that in the low interspecies interaction regime, the tunneling process is dominated by the A bosons and occurs between the left and right wells, while the B boson is strongly localized.

\begin{figure}
\includegraphics[width=16cm]{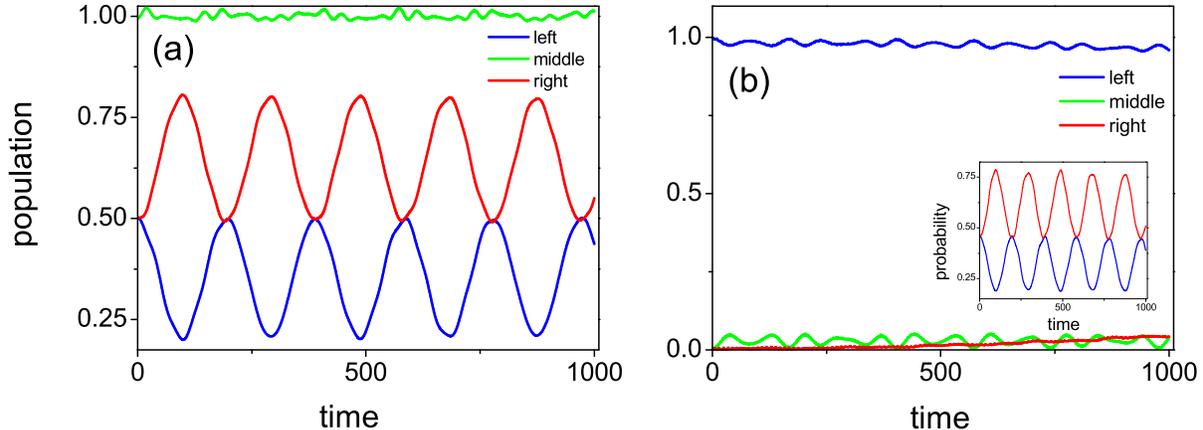}
\caption{The population oscillations of a mixture composing of two A bosons and one B boson with the initial state $(|1,1,0\rangle_{A}+|0,1,1\rangle_{A})|1,0,0\rangle_{B}$. (a) Population of A bosons in the TPH well, where the oscillation is mainly between the right and left wells. (b) Population of the B boson, which remains localized in the left well. The calculations are done via MCTDH, with the interaction strengths $g_{A}=0.5$ and $g_{AB}=0.05$. The insert (figure 3(b)) shows the probability of the number states $|1,1,0\rangle_{A}|1,0,0\rangle_{B}$ and $|0,1,1\rangle_{A}|1,0,0\rangle_{B}$, and the tunneling is mainly between these two number states. } \label{figure 3}
\end{figure}

In order to explain the numerical results calculated by the $ab-initio$ MCTDH method, we again rely on a number-state analysis. The initial state is essentially the superposition of two number states $|\varPsi(t=0)\rangle=|1,1,0\rangle_{A}|1,0,0\rangle_{B}+|0,1,1\rangle_{A}|1,0,0\rangle_{B}$, named in the following as $|a\rangle$ and $|b\rangle$, respectively. The spatial correlation of the A species between the left and right wells is reflected by this superposition in terms of number states. The time evolution of the system is the linear superposition of two tunneling branches, one of which takes $|a\rangle$ as initial state, and the other one takes $|b\rangle$ as the initial state. The tunneling properties in terms of the expectation value of an operator with the state $|\varPsi\rangle=|\varPsi(t)\rangle$ can be expressed as:

\begin{equation}
 \langle\varPsi|\hat{L}|\varPsi\rangle=\sum_{\alpha,\beta}\langle \alpha|\hat{L}|\beta\rangle (\langle \beta|a\rangle \langle a|\alpha\rangle +\langle \beta|b\rangle \langle b|\alpha\rangle+2\langle \beta|b\rangle\langle a|\alpha\rangle) e^{i(e_{\beta}-e_{\alpha})t},
\end{equation}
where $\alpha$,$\beta$ label the eigenstates of the complete interacting mixture, and $e_{\alpha}$ is the eigenenergy of the $\alpha$ eigenstate. The first two terms in the expectation value are the separate contribution from the two branches, and the third term comes from the interference of the two branches. The expectation value of $\hat{L}$ contains the separate contribution from both tunneling branches, as well as the interference of the two branches. The interference term comes from the superposition of $|a\rangle$ and $|b\rangle$ in the initial state, and it is equivalent to saying that the quantum interference is induced by the spatial correlations.

In the low interspecies interaction regime, only the two number states $|a\rangle$ and $|b\rangle$ contribute significantly to the tunneling process (as shown in the inset of figure 3(b), which provide the time-evolution of the probability of the occupation of these two number states), and we can therefore employ the reduced two-state basis consisting of $|a\rangle$ and $|b\rangle$, while the two eigenstates in this reduced basis for $g_{AB}=0$ are $|a\rangle\pm|b\rangle$. A direct calculation shows that the first two terms in equation (3) cancel each other, and the tunneling properties, such as the population oscillation in figure 3(a), results exclusively from the quantum interference of the two branches. The latter is induced by the spatial correlations, and we call this behavior therefore correlation-induced tunneling. Such a correlation-induced tunneling strongly depends on the coherent correlations between the left and right wells, and could work as a measure for the coherence between the two wells. In order to illustrate this point, let us suppose that the TPH well is coupled to some decoherent source, for instance the left well is connected to a classical environment, which causes a randomly temporally fluctuating phase difference between the left and right wells. The random phase difference is then mapped to that between the states $|a\rangle$ and $|b\rangle$. The interference is destroyed by this random phase difference due to decoherence between the left and right wells, and the correlation-induced tunneling is consequently destroyed. In this sense, the destruction of correlation-induced tunneling can measure the coherence between the left and right wells.

Another interesting point is to examine the correlation-induced tunneling process from the perspective of energy conservation. A common mechanism for enhanced tunneling processes occurs if the tunneling takes place between states which are energetically in resonance meaning that the onsite energies consisting of the sum of potential and interaction energies become equal. However, the correlation-induced tunneling starts when the two number states $|a\rangle$ and $|b\rangle$ are detuned and consequently $|a\rangle\pm|b\rangle$ become non-degenerate by the onset of interspecies interaction. The A boson then tunnels from the left well of higher interaction energy to the right well with lower interaction energy, which decreases the total interaction energy, and breaks the condition for resonant tunneling. The underlying mechanism is the following: the spatial correlation stores some portion of the total energy in the form of kinetic energy, and during tunneling, on the one hand the interaction energy of the system decreases, while on the other hand, the kinetic energy increases in response to the change of correlation of the time-evolved wave function. The loss of interaction energy is compensated by the gain of kinetic energy.

\begin{figure}
\includegraphics[width=16cm]{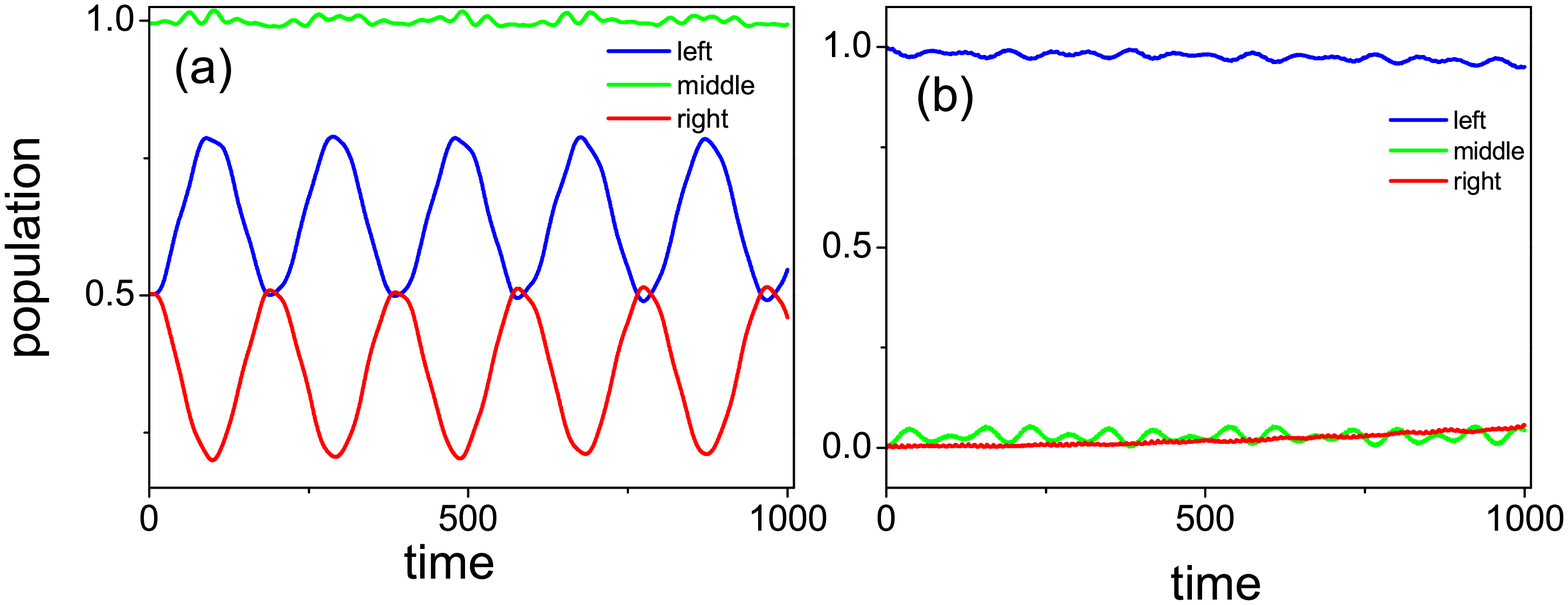}
\caption{The population oscillations of a mixture system initially prepared in the dark state $(|1,1,0\rangle_{A}-|0,1,1\rangle_{A})|1,0,0\rangle_{B}$. (a) Population oscillation of A bosons in the TPH well. (b) Population oscillation of the B boson. Parameters are the same as in figure 3.} \label{figure 4}
\end{figure}

In order to estimate the gain of kinetic energy, we calculate the kinetic energy of the initial state $|a\rangle+|b\rangle$ as $2J_{eff}$. $J_{eff}$ is the effective coupling between $|a\rangle$ and $|b\rangle$, which is mediated through the state $|0,2,0\rangle_{A}|1,0,0\rangle_{B}$, and can be estimated as $J_{eff}=J^{2}/\Delta E$, where $J$ is the hopping strength of A bosons between neighboring wells in the system, and $\Delta E=\epsilon_{|1,1,0\rangle_{A}|1,0,0\rangle_{B}}-\epsilon_{|0,2,0\rangle_{A}|1,0,0\rangle_{B}}$ is the energy difference between the initial state and the intermediate state. In our case, $\Delta E$ is less than zero, and $J_{eff}$, and the associated kinetic energy of the initial state, is negative. During tunneling, the correlation encoded in the superposition of the number states decreases, and the kinetic energy goes from the negative value of $2J_{eff}$ to zero, which compensates for the loss of interaction energy. From this point of view, we see that the correlation-induced tunneling process brings the kinetic energy into play, and goes beyond the picture of resonant tunneling which is based exclusively on the comparison of the onsite energies.

From the viewpoint of energy conservation, we can also estimate the regime of interspecies interaction strength $g_{AB}$ for which the correlation-induced tunneling processes occurs. To compensate the loss of the interaction energy by the kinetic energy, the detuning of the interaction energy which is approximately $g_{AB}$ must be of the same order as the kinetic energy of the initial state, $2J_{eff}$, $i.e.$, the regime of $g_{AB}$ for correlation-induced tunneling should be of the order of $2J_{eff}$.

Following the line of argumentation of energy conservation, when the initial state is prepared as $|a\rangle-|b\rangle$, the so called dark state, where the correlation $G_{LR}$ becomes negative, and the kinetic energy of the initial state becomes positive, we should observe the counterintuitive phenomenon that A bosons flow upstream from the weak-interaction right well to the strong-interaction $(g_{AB})$ left well. Indeed, we do observe such an upstream flowing, when we prepare the initial state $|a\rangle-|b\rangle$, as shown in figure 4. The profile of the evolution of the population is similar to that shown in figure 3, where the initial state is prepared as $|a\rangle+|b\rangle$, but the tunneling direction is now from the right well to the left well, opposite to that in figure 3.

 From figure 3 and 4, it becomes clear that the onset of the interspecies interaction $g_{AB}$ disturbes the spatial correlation of A species, and this leads to population oscillations of the A bosons between the sites which are correlated in the wavefunction, that is the left and right wells of our system. The population of the uncorrelated site, $i.e.$ the middle well is barely affected. Moreover, the correlation does not only determine which sites participate in the tunneling, but also the direction of the tunneling, which offers the possibility to control the tunneling processes by spatial correlations.

As a final remark, it is also worth mentioning that the correlation-induced tunneling is triggered by the disturbance of the correlation of the system, and this is not restricted to a few-body system, since spatial correlations exist also in many-body systems. We could also expect to observe such phenomena in many body system, for instance in a double well. If the condensates in the two wells present correlations, a weak tilt of the system, which detunes the potential energy of condensates in different wells can also result in such a correlation-induced tunneling between the left and right wells, and the amplitude of the tunneling can be used to measure the strength of the correlation between the left and right wells, in other words, the coherence of the neighboring wells. In this way the correlation-induced tunneling manifests itself as an alternative method to measure the coherence of two separate condensates, besides the interference fringes after overlapping of the condensates.

\subsection{Correlated tunneling in the intermediate regime}

As we increase the interspecies interaction beyond the weak interaction regime, of which the order of magnitude has been estimated in the previous section, the two number states contributing to the initial state become increasingly different with respect to their energies and they do not contribute significantly to the same eigenstates, thereby destroying the interference of the two tunneling branches. In this sense the correlation-induced tunneling process disappears. The tunneling evolution in terms of the expectation value, as shown in equation (3), is then the direct summation of each branch separately without interference term.

Generally speaking, the number states $|1,1,0\rangle_{A}|1,0,0\rangle_{B}$ and $|0,1,1\rangle_{A}|1,0,0\rangle_{B}$ are energetically in resonance with their spatially-symmetric counterparts $|0,1,1\rangle_{A}|0,0,1\rangle_{B}$ and $|1,1,0\rangle_{A}|0,0,1\rangle_{B}$, respectively, almost everywhere in the intermediate regime except for narrow resonant windows, where some other number states become resonant with one of the initial number states. Out of these narrow resonant windows, the tunneling of the system is practically delayed, due to the extremely weak coupling between the initial number states with their spatial-symmetric counterparts. In the resonant windows, however, correlated tunneling between A and B species are observed, and we show an example therefore in figure 5.

\begin{figure}
\includegraphics[width=16cm]{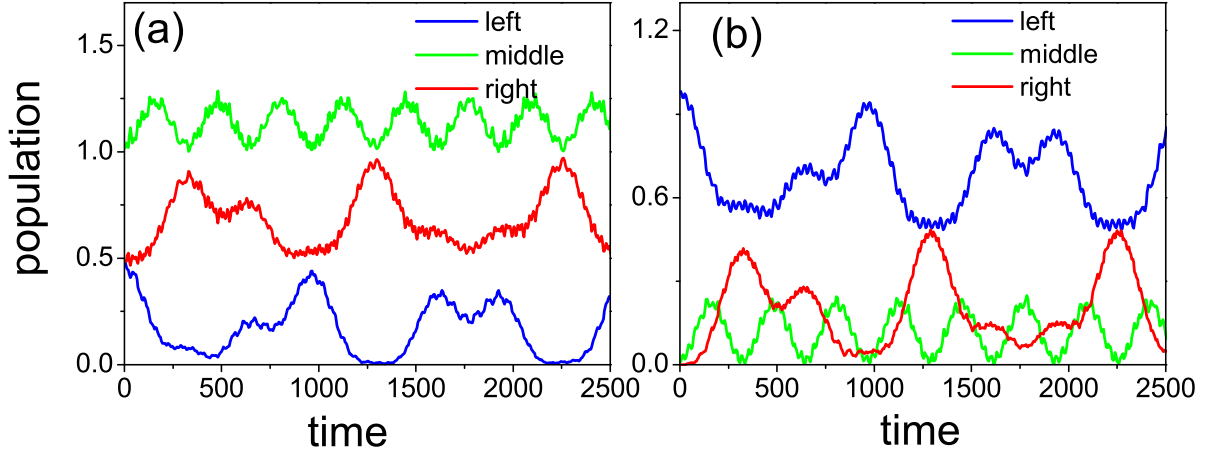}
\caption{The population oscillation of (a) the A species, (b) the B species in the TPH well, with the initial state $(|1,1,0\rangle_{A}+|0,1,1\rangle_{A})|1,0,0\rangle_{B}$, and for the interspecies interaction strength $g_{AB}=0.13$ in the intermediate regime. From the density oscillations we observe correlated tunneling of A bosons and the B boson.} \label{figure 5}
\end{figure}

In figure 5, $g_{AB}$ resides in the resonant window, where $|1,1,0\rangle_{A}|1,0,0\rangle_{B}$ is in resonance with $|0,2,0\rangle_{A}|0,1,0\rangle_{B}$. Figure 5 (a) and (b) show the population oscillation of the A and B species, respectively. We can see that the population profiles of the A and B species in the same well match each other, which indicates the correlated tunneling behavior. Meanwhile, the populations of the A boson in the left and right wells oscillate in the intervals $[1/2,0]$ and $[1/2,1]$, respectively, and the populations of the B boson in the left and right wells oscillate in the intervals $[1,1/2]$ and $[0,1/2]$, respectively. Therefore, half of the population of A bosons and B bosons tunnels from the left to the right wells simultaneously, and we therefore encounters a simultaneous tunneling of 'half' A boson and 'half' B boson. After half a period of the tunneling, the correlation of the A bosons between the left and right wells vanishes, as now the A bosons initially residing in the left and right wells have tunneled to the right well completely. On the other hand the B boson initially completely localized in the left well now splits its probability into half in the left well and half in the right well, and as a consequence, we observe a spatial correlation of the B species between the left and right wells. In other words, in the course the tunneling, the spatial correlations are transferred between the A and B species.

Via the above example, we have demonstrated a certain mechanism of correlated tunneling in mixtures. During the tunneling process, the spatial correlations between the left and right wells are transferred forth and back between the two species, which could be used as dynamical writing and reading of quantum coherence between qubits, as well as gate operation of qubits, and could potentially find applications in quantum computing.

\subsection{Higher band contributions in the strong interspecies interaction regime}

Up to this point, we have focused on the situation where $g_{AB}$ is not large enough to excite higher band states during the onset of $g_{AB}$, and the main effect of the onset of $g_{AB}$ is to modifying the onsite interaction of related number states. In this section we discuss the tunneling mechanisms, when $g_{AB}$ is large enough such that the higher band states get into play, and this cannot be covered by a standard single-band approximation.

In the strong interspecies interaction regime, the two tunneling branches with initial states $|1,1,0\rangle_{A}|1,0,0\rangle_{B}$ and $|0,1,1\rangle_{A}|1,0,0\rangle_{B}$, respectively, are not coupled to each other, and they evolve separately. We can obtain the tunneling properties of the system by separately studying the tunneling of each branch. For the tunneling branch with the initial state $|0,1,1\rangle_{A}|1,0,0\rangle_{B}$, where the A and B boson reside in different wells initially, and they cannot interact with each other, the onset of $g_{AB}$ will not affect this initial state, and the tunneling of this branch will remain practically self-trapped, due to a large onsite energy detuning.

\begin{figure}
\includegraphics[width=16cm]{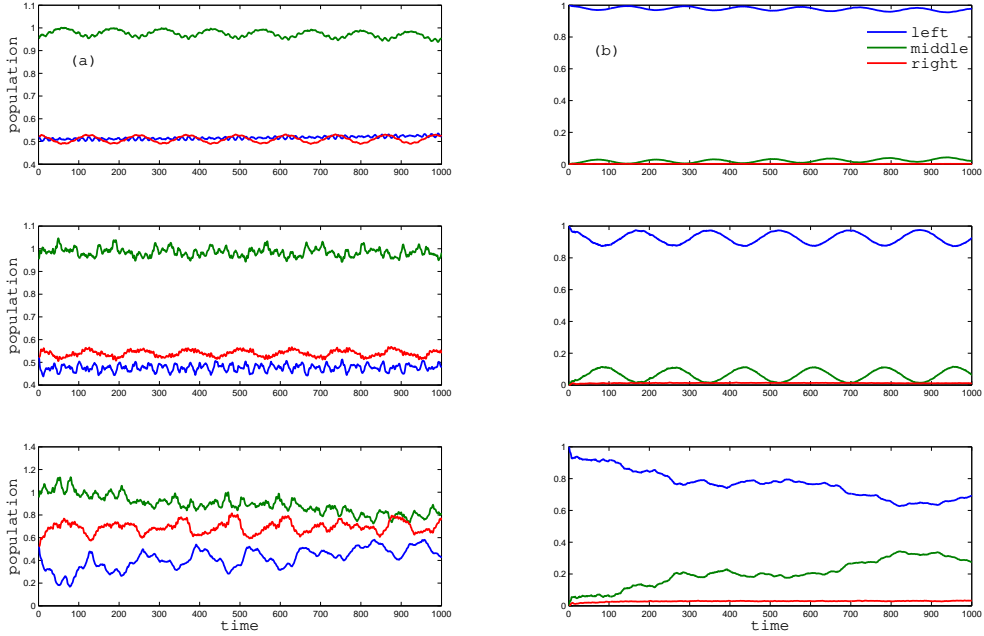}
\caption{Tunneling behavior in the strong interspecies interaction regime: population oscillations of (a)-column the A species, (b)-column the B species, with the initial state $(|1,1,0\rangle_{A}+|0,1,1\rangle_{A})|1,0,0\rangle_{B}$. In both columns, the interspecies interaction increases from top to bottom, taking values of $g_{AB}=1.0$, $g_{AB}=3.0$, and $g_{AB}=7.0$, with $g_{A}=9.0$. From the density oscillations we observe enhanced tunneling as $g_{AB}$ increases. } \label{figure 6}
\end{figure}

On the other hand, an instantaneous onset of $g_{AB}$ will strongly affect the initial state of $|1,1,0\rangle_{A}|1,0,0\rangle_{B}$, and higher band states could be excited, as the B boson is in the same well together with one A-boson. The higher band states have a larger hopping strength between neighboring wells, as the wavefunction of the higher band states can penetrate further into the barrier and have a larger overlap among states in neighboring sites \cite{highband-wf}. This can lead to enhanced tunneling. In figure 6, we show the tunneling dynamics of A and B bosons for different interspecies interaction strength $g_{AB}$. We observe that tunneling is enhanced with increasing the interaction strength, which is illustrated by the increase of the tunneling amplitude, in spirit of the detuning of the onsite energy. 

\begin{figure}
\includegraphics[width=16cm]{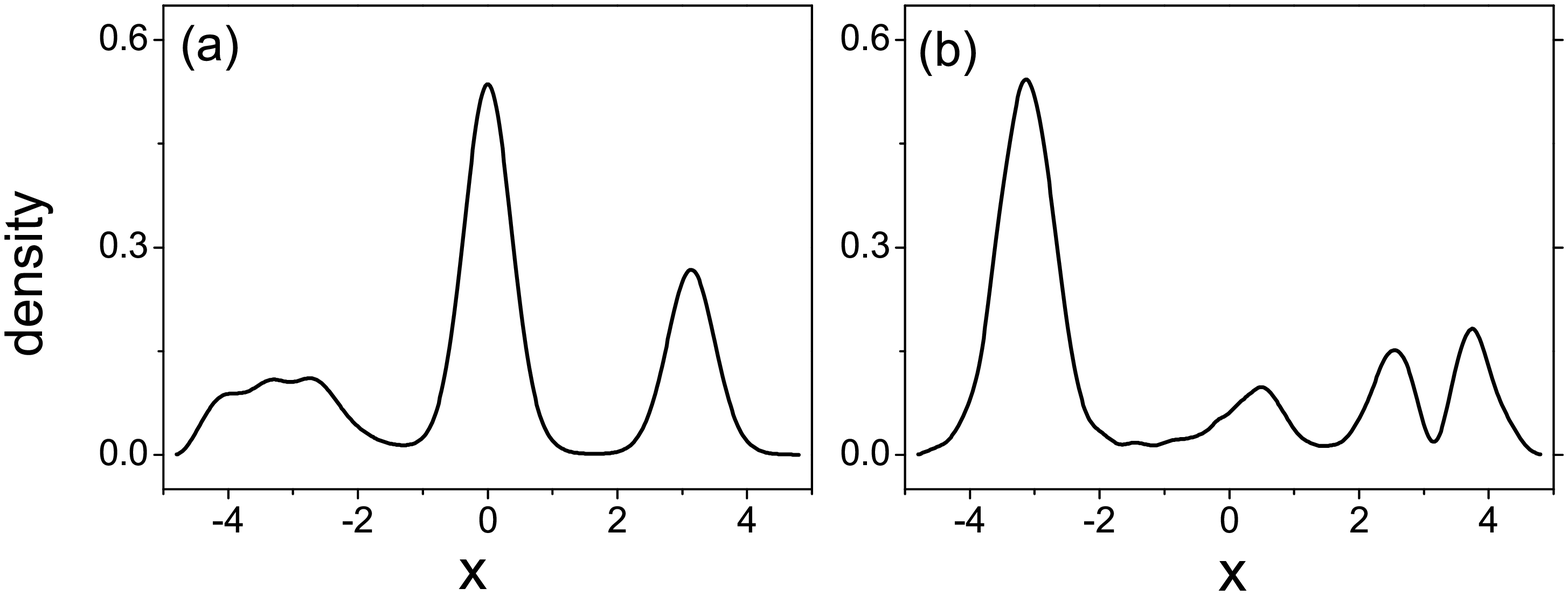}
\caption{One-body density profile of (a) A-species bosons and (b) B-species boson at a time instant during tunneling. In (a) we observe the reshaping of the density from a Gaussian, which indicates a combination of ground and excited levels in this well. In (b), the ground level is locating in the left and middle well, while the excited portion tunnels to the right well, where we see a nodal profile. The one-body density profiles are taken from the propagation with $g_{A}=9.0$ and $g_{AB}=7.0$.} \label{figure 7}
\end{figure}

In figure 7 we show the one-body density profile of the A bosons and B boson at a certain time instant during tunneling in order to visualize the higher band excitation. Figure 7(a) shows the one-body density profile of the A bosons, and we observe the reformation of the wavepacket in the left well. Initially the A bosons occupy the ground level of all three wells, and are only partially excited in the left well by the onset of $g_{AB}$ resultingly, we observe the density of a superposition of the ground and excited states, which blurs the signature of a higher band excitation. The B boson, however, is localized in the left well initially, and we observe a clear signature of a higher band excitation, when tunneling to a higher band state in the right well occurs. In figure 7(b), we observe the density profile of a ground state in the left and middle well while the first excited state is populated in the right well. The enhanced tunneling therefore works like a filter, which separates the ground and excited states into different wells.

\section{Summary and Conclusion}
In this work we study mechanism of the tunneling dynamics of few-boson mixtures confined to a combined triple well and harmonic trap. The harmonic trap induces spatial correlations to one species, and we explore how such correlations affect the tunneling properties of the mixture. The tunneling dynamics is qualitatively different for different the interspecies interaction regimes. In the weak interspecies interaction regime, we observe a so-called correlation-induced tunneling, and in the intermediate regime, correlated tunneling between different species is observed whereas in the strong interaction regime, higher band states come into play, and tunneling is enhanced due to an effective weakening of the optical lattice barriers.

Our main result is the correlation-induced tunneling in the weak interspecies interaction regime. Such tunneling arises due to exposure of spatial correlations to weak perturbation, and is a novel mechanisms different from the common resonant tunneling process. We demonstrate that the spatial correlations can well control the dynamics of the correlation-induced tunneling, including the participating wells and the tunneling direction. We also discuss possible applications of the correlation tunneling, such as the measure of off-site coherence, and the generalization to many-body systems.

In the intermediate interspecies interaction regime, the correlation-induced tunneling effect disappears, as the high energy detuning between the involved states contributing to the correlation destroys the interference of different tunneling branches. The tunneling of both species is now delayed except in some resonant windows of the interaction strength. For the latter we observe correlated tunneling of both species, which can dynamically transfer the correlation initially present in one species to the other species. We propose that such correlated tunneling can work as writing and reading and gate operations between qubits.

Finally in the strong interspecies interaction regime, higher band states are excited in the course of the onset of interspecies interaction, and we observe enhanced tunneling via higher band states, where the tunneling amplitude increases with increasing interaction strength.

\begin{acknowledgments}
L. C. and P. S. gratefully acknowledge financial support by the Deutsche Forschungsgemeinschaft (DFG), and B. C. acknowledges support from the Landesexzellen- zinitiative Hamburg, which is financed by the Science and Research Foundation Hamburg and supported by the Joachim Herz Stiftung..
\end{acknowledgments}

\bibliographystyle{apsrev4-1}
\bibliography{reference}

\end{document}